# Optimization of Residential Demand Response Program Cost with Consideration for Occupants Thermal Comfort and Privacy

Reza Nematirad, Student Member, IEEE, M. M. Ardehali, and Amir Khorsandi

*Abstract*--Residential consumers can use the demand response program (DRP) if they can utilize the home energy management system (HEMS), which reduces consumer costs by automatically adjusting air conditioning (AC) setpoints and shifting some appliances to off-peak hours. If HEMS knows occupancy status, consumers can gain more economic benefits and thermal comfort. However, for the building occupancy status, direct sensing is costly, inaccurate, and intrusive for residents. So, forecasting algorithms could serve as an effective alternative. The goal of this study is to present a non-intrusive, accurate, and cost-effective approach, to develop a multi-objective simulation model for the application of DRPs in a smart residential house, where (a) electrical load demand reduction, (b) adjustment in thermal comfort (AC) temperature setpoints, and (c) , worst cases scenario approach is very conservative. Because that is unlikely all uncertain parameters take their worst values at all times. So, the flexible robust counterpart optimization along with uncertainty budgets is developed to consider uncertainty realistically. Simulated results indicate that considering uncertainty increases the costs by 36 percent and decreases the AC temperature setpoints. Besides, using DRPs reduces demand by shifting some appliance operations to off-peak hours and lowers costs by 13.2 percent.

*Index Terms*-- Demand response program, Thermal comfort, Occupant privacy, Occupancy forecasting, Random forest, LightGBM, MLP-ANN, Robust counterpart optimization, Multi-objective genetic algorithm.

NOMENCLATURE

**Identifiers**

| | |
|---|---|
| $h_i$ | Index of time, from 1 to H |
| $h_s^{start}$ | Start of next interval |
| $h_s^{end}$ | End of a period |
| $k$ | Index used to represent time lags in the ARX model |
| $S$ | Index used for demand shiftable appliances at the residential building |
| $I$ | Index used for the nodes in the community circuit |
| $J$ | Index used for the load points (customers) connected to each node i |

**Variables and Constants**

| | |
|---|---|
| A | Coefficient matrix of linear programming |
| $\gamma^{occ}$ | Occupancy uncertainty budget |
| $\gamma^d$ | Demand uncertainty budget |
| b | Right side vector of inequality |
| c | Linear programming vector |
| d | Linear programming vector |
| $c_h^u$ | Cost of electricity at time h ($/kW) |
| $c_c$ | Penalty or reward price at time h($/kW) |
| L | Number of uncertain data |
| N | Set of nodes in the community circuit |
| M | Set of exogenous inputs to the ARX model (i.e. outdoor temperature, occupancy, and AC unit temperature setpoints) |
| O | Objective function |
| $O_{cc}$ | Occupancy level |
| $O^{Ro}$ | Uncertain objective function |
| $p_h^{AC}$ | Consumers demand (kW) |
| $p_h^d$ | Consumers demand (kW) |
| $p_h^{shift}$ | Shiftable loads (kW) |
| $p_h^{mis}$ | Miscellaneous electrical loads (kW) |
| $p_h^{d,des}$ | Desired demand at time h(kW) |
| $p_h^{d'}$ | Internal demand except for air conditioning power(kW) |
| $P_h^u$ | Purchased power from the utility (kW) |
| $p_{i,j,h}^{AC}$ | Air conditioning temperature setpoints (°C) |
| $T^{AC,set}$ | Air conditioning temperature setpoints (°C) |
| $T^{AC,des}$ | Air conditioning desired temperature (°C) |
| $u_{s,h}$ | Binary variable indicating if the shiftable load *s* is ON at time *t* (= 1: ON, 0: OFF) |
| W | Auxiliary variable |
| W' | Auxiliary variable |
| X | linear programming vector |
| Z | Auxiliary variable |
| Z' | Auxiliary variable |
| U | Set of uncertain parameters |

**Greek symbols**

| | |
|---|---|
| $\gamma$ | Uncertainty budget |
| $\gamma'$ | Uncertainty budget |
| Ψ | Set of uncertainty |
| ς | Uncertainty adjustment factor |
| α | Coefficient of the ARX model |
| β | Coefficient of the ARX model |
| Γ | Auxiliary variable |

## I. INTRODUCTION

Electricity demand has experienced a noticeable increasing growth over the past decades that has resulted in a wider supply-demand gap and utilities have introduced demand response programs (DRP) to address these concerns [1]. The ultimate goal of utilities is to provide customers with opportunities to reduce their demands and electricity costs by means of either incentive-based or time-based rate schemes [2]. In an incentive-based DRP, customers participate voluntarily in specific rewarding programs and allow the operators to control some of their demands from major electric appliances [3]. However, it is difficult for consumers to manually manage their appliance operation scheme, and a home energy management system (HEMS) is essential to take advantage of the DRPs more efficiently to reduce demand and cost [4]. However, the consumers comfort cannot be compromised [5]. Accordingly, the comfort level of consumers must be accounted for by the HEMS. Finally, to evaluate the problem realistically, the inherent uncertainty in the associated parameters must also be considered, which causes the problem to be more complicated. The literature review on applications of DRPs with considerations for occupant thermal comfort indicates a few studies focus on non-intrusive, accurate, and cost-effective approaches to forecast occupancy levels and evaluate the applications of DRPs with consideration for the uncertainty of data, simultenously.

### A. Literature review

#### 1. Demand response and home energy management system

Recently several studies have examined the implementation of DRPs in smart residential buildings and houses from economic, technical, and behavioral perspectives. DRPs require consumers to manage and monitor their power consumption, which is unlikely to be accomplished manually [6]. So, HEMSs have been developed for implementing of DRPs into smart buildings successfully [7]. HEMS is responsible for scheduling or shifting the consumption of some appliances into off-peak hours in order to reduce the consumer energy cost [8] and adjust the air conditioning (AC) temperature setpoints corresponding with DRP schemes [9].

A multi-objective optimization is proposed in [10] considering demand response management and thermal comfort in a microgrid. Here, the cost of energy and thermal comfort is considered as the objective functions. The results demonstrate an improvement of nearly 12% in costs, where precooling schemes are used for ensuring consumer comfort. However, in that study, occupancy forecasting, uncertainty, and demand reduction are not considered. Another study proposes a multi-objective optimization methodology in individual neighborhoods to maximize consumer comfort, reduce demand, and minimize battery operational costs [11]. In that study, HEMS is used for modeling and management of shiftable loads. However, occupancy detection, peak reduction cost, and the inherent uncertainty in the parameters are not addressed. Further, HMES is used to manage the operation time of the washing machine and dryer which can be shifted to lunch or dinner time [12]. In that study, the problem is formulated as a multi-objective function, and simulation using the cooperative particle swarm optimization algorithm shows an improvement in cost by %6 by applying DRPs. Nevertheless, the effect of occupancy estimation and the inherent uncertainties are not considered.

Besides, to analyze the participation of customers in DRPs, a load model is necessary [13]. For example, physical load models in residential DRP applications are proposed in [14]. In reference [15], the estimation of a one-day-ahead household load system for the HEMS appliances is addressed by introducing a feedback-type error-correcting function using artificial intelligence.

#### 2. Occupancy forecasting

Despite the many benefits of using HEMS for implementing DRPs, challenges related to occupancy forecasting exist. As the effectiveness of DRPs is heavily influenced by building occupancy levels [4], for thermal comfort in residential buildings, the presence and absence of residents can substantially affect the AC temperature setpoints [16]. There are several technologies and devices such as video cameras, infrared cameras, passive infrared sensors, and break-beam sensors that can be used to detect occupancy levels. For example, a single passive infrared sensor is applied in [17] to estimate the number of occupants in a residential home. In [18], active ultrasonic sensing is used to estimate the number of people in a space. To better monitor the occupancy changes cloud-based technologies are used [19] Also, the occupancy changes can be estimated by analyzing the dynamic $CO_2$ concentration [20]. As proved in [21], Wi-Fi connection data results are more accurate than carbon dioxide sensors to estimate occupancy levels. While these technologies and devices that directly monitor occupancy levels can provide valuable data for building energy management systems, they have several drawbacks. For example, building occupants privacy could be potentially violated or they could be misused by irrelevant entities. Additionally, such technologies can be expensive to install and maintain. They are limited and can only estimate the current occupancy state as well [22]. Alternatively, statistical models that employ historical data are more appropriate than direct monitoring and sensing since they can reduce the cost and risk of privacy violation, and forecast occupancy levels for the next hours [23].

The effectiveness of statistical and machine learning techniques has been demonstrated in several power system applications such as electric vehicle coordination [24], fault type detection in distribution transformers [25], and solar irradiance forecasting [26]. However, a few studies have applied these approaches to occupancy detection. For example, the authors in [27] propose a statistical method for occupancy detection that estimates occupancy levels based on historical data of consumer demand, and according to the results, the algorithm is accurate, especially during the summer, where there is a strong correlation between occupancy levels and electricity demand. However, that study needs a few smart meters and can only detect the current occupancy status. Also, traditional machine learning algorithms, such as support vector machines and K nearest neighbors, are used in [28] to classify the occupants levels, and the results indicate that K nearest neighbors outperform. However, the mentioned algorithms in that study, are not able to deal with imbalanced data [29]. Therefore, powerful techniques, such as the next generation of machine learning algorithms that are able to work with imbalanced data and provide a non-intrusive, accurate, and cost-effective solution to forecast occupancy levels for the application of DRPs are required.

### 3. Occupants thermal comfort

While HEMS seeks to adjust AC temperature setpoints to reduce the demand during peak hours, occupants thermal comfort must not be neglected. Some studies have integrated thermal comfort into the formulation of optimization problems [30]. For instance, a predictive model controller is proposed in [31] to optimize residential AC energy consumption considering occupants thermal comfort, and simulation results show that peak demand can be reduced by 5%, while occupants thermal comfort is maintained. A practical method is proposed in [32] for calculating and managing occupants preferences to maximize energy efficiency. Also, in [33], the effects of different AC temperature setpoints on occupants thermal comfort and energy efficiency are reviewed. Several advanced control strategies are proposed in [34], where one adjusts the air supply rate to maintain indoor air quality, and another evaluates thermal comfort. In that study, the simulations show that the proposed control strategies can reduce total energy costs by 11%. Although these studies have integrated occupants thermal comfort in their problem, there is still the lack of a comprehensive approach to consider occupancy estimation and uncertainty on the related data to take advantage of DRPs more efficiently.

### B. Contributions

Based on the literature of the study, the main contributions of this study are as follows.
- Utilization of next-generation of machine learning algorithms that provides a non-intrusive, accurate, and cost-effective solution to forecast occupancy levels for the application of DRPs.
- Integrating electrical load demand reduction, the adjustment in thermal comfort AC temperature setpoints, and minimizing consumer cost into a multi-objective optimization problem.
- Developing robust counterpart optimization along with different uncertainty budgets to evaluate the uncertainty in the data.
- Evaluating the applications of DRPs with consideration for uncertainty in the data, simultaneously.

The problem is formulated as a multi-objective model to meet the requirements for electricity demand reduction, the adjustment in thermal comfort AC temperature setpoints, and minimizing consumer cost subject to the related constraints. In addition, the robust counterpart optimization technique along with uncertainty budgets are employed to help HEMS to consider the inherent uncertainty in the forecasted data. In order to assess the impact of applying DRPs and uncertainty on the results, the multi-objective model is simulated for four case studies where all possible combinations of applying and not applying DRPs and uncertainty are examined.

The remaining manuscript is organized as follows. Section 2 discusses machine learning algorithms and occupancy data estimation. The related mathematical models are given in Section 3. In Section 4, the multi-objective function is formulated deterministically, then the robust counterpart optimization is applied to take uncertainty into account. Discussions and explanations of the numerical results are provided in Section 5. Finally, in Section 6, recommendations and conclusions are provided.

## II. OCCUPANCY DATA ESTIMATION

### A. Data preprocessing and preparation

To forecast occupancy levels based on machine learning algorithms, two sets of initial datasets as inputs (or features) and outputs (or responses) are required [35]. This study uses hourly electricity demand data as features and occupancy level data as responses for machine learning algorithms. For hourly data of the occupancy levels, Wi-Fi connection records are used and occupancy data are collected. Finally, these sets of raw data are converted into acceptable inputs and outputs for forecasting.

### B. Forecasting algorithms

In this work, three machine learning algorithms including random forest [36], light gradient boosting machine (lightGBM) [36], and multilayer perceptron artificial neural network (MLP-ANN) [37] are used to forecast occupancy levels for using by HMES. Random forest and LightGBM are ensemble algorithms that use multiple decision trees to make a prediction. The random forest and LightGBM algorithms are resistant to overfitting, have inherent feature selection, and are highly accurate and efficient, but may not be suitable for very high-dimensional data sets [36]. ANN-MLP is a deep learning algorithm that can learn complex patterns in the data that is appropriate for working with high-dimensional data and complex models. However, overfitting, computationally intensive, and significant tuning and optimization of hyperparameters are its disadvantages [37]. The use of multiple algorithms can assist in identifying the underlying patterns in the data. By comparing the results of different algorithms, it is possible to get insight into how they perform on the same data and achieve better accuracy. Supervised machine learning algorithms include regression and classification processes, where they are used for prediction and work with labeled datasets [29]. Regression is necessary in this study because (a) many features with continuous values are lost when the outputs are defined as a class (discrete values) in the classification process, and classification accuracy may be reduced, (b) if the input data distribution is heavily imbalanced, the classification result is non-uniform, and in practice, all outputs could shift to the one class [29]. To determine the accuracy of the machine learning algorithms, root mean square error (RMSE), mean

## III. MATHEMATICAL MODELING

### A. Loads

For the purpose of this study, the internal load demand of a residential house is separated into four categories including AC, shiftable, non-shiftable, and miscellaneous loads [18]. Shiftable loads such as dishwashers, washers and dryers can be transferred to a different time slot and operate according to their patterns. Appliances including ovens, TV, and personal computers are considered as non-shiftable loads [39]. To forecast the current cooling load, the methodology presented in [11] is employed, where a black-box reduced order model called auto-regressive with three exogenous inputs (outdoor temperature, occupancy, and AC temperature setpoints) and a previous cooling load is used to predict the current AC loading given by [11],

$$p_{i,j,h}^{AC} = \sum_{k \in K}\left[-\alpha_{i,j,k} \times p_{i,j,(h-k)}^{AC} + \sum_{m \in M}\beta_{i,j,k,m} \times Occ_{i,j,m,(h-k+1)}\right] \quad (1)$$

$$\forall_i \in N, \forall_j \in D, \forall_h \in H$$

since this study is conducted just for one smart home, M and j are set to 1. As a result, load demand can be expressed as follows,

$$p_h^d = p_h^{shift} + p_h^{n-shift} + p_h^{mis} + p_h^{AC}, \forall_h \in H \quad (2)$$

### B. Robust counterpart optimization

Traditionally, optimization problems are solved by assuming data is deterministic, although most data in real life is uncertain. In real-world problems, changing one of the data points can violate a number of constraints (assuming the problems are addressed deterministically), which results in a non-optimal or even impossible solution. The primary idea behind dealing with uncertainty is to evaluate the worst-case scenario and conduct robust counterpart optimization based on it. However, in the absence of uncertainty budgets, the worst-case scenario approach may lead to very conservative solutions. Because all uncertain data rarely hold their worst values at the same time. In optimization problems, uncertainty budgets are used to quantify the uncertainty in the data and to determine what level of uncertainty is acceptable (not simply worst values). Therefore, robust counterpart optimization with uncertainty budgets is utilized in this study to account for uncertainties and enable to develop more realistic and effective solutions [40].

*1. Robust counterpart*

Consider the optimization problem given by,

$$O = \{c^T.x + d\} \quad (3)$$

s.t:

$$A.x \leq b \quad (4)$$

If coefficient matrices c, d, A, b are uncertain and belong to uncertainty set U, Eq. (3) can be written as [40],

$$O^{Ro} = \left\{\min_x\{c^T.x + d : A.x \leq b\}\right\}_{(A,b,c,d)\in U} \quad (5)$$

We suppose that the uncertainty set is affinely parameterized by a perturbation vector $\varsigma$ altering in a perturbation set Z. For instance, the parameter c with uncertainty can be written as [40],

$$\tilde{c} = c^0 + \sum_{q \in Q}\varsigma_q \times \hat{c}_q : \varsigma \in Z \subset R^Q \quad (6)$$

$$Z = \{\varsigma \in R^L : |\varsigma| \leq 1,\}$$

then the optimization problem with assuming the worst-case scenario for all uncertain variables is as follows [40],

$$O^{Ro} = \min_x\left\{\max_{(A,b,c,d)}\left[c^T.x + d\right], A.x \leq b, (A,b,c,d) \in U\right\} \quad (7)$$

*2. Budget box*

While it is nearly unlikely that all uncertain parameters coincide, by allocating budgets to uncertain parameters, the deviation from their nominal value can be controlled and limited. So, by taking budgets into account, perturbation set Z is rewritten as [40],

$$b = b^0 + \sum_{l=1}^{L}\varsigma_l[b^l] \quad (8)$$

where L indicates the number of uncertain parameters. To develop robust counterpart optimization in the general case, the robust counterparts of the objective functions and constraints must be obtained. When A and b are uncertain, according to Eq. (6) they can be rewritten as follows [40],

$$A = A^0 + \sum_{l=1}^{L}\varsigma_l[A^l] \quad (9)$$

$$b = b^0 + \sum_{l=1}^{L}\varsigma_l[b^l] \quad (10)$$

Therefore, by using Eqs. (9) and (10), Eq. (4) is written as follows,
According to [41], the left-hand side of Eq. (12) can be written

$$\left[A^0\right]^T x + \sum_{l=1}^{L}\varsigma_l\left[A^l\right]^T x \leq \left[b^0\right]^T + \sum_{l=1}^{L}\varsigma_l\left[b^l\right]^T \quad (11)$$

$$\sum_{l=1}^{L}\varsigma_l\left(\left[A^l\right]^T x - \left[b^l\right]\right) \leq \left[b^0\right] - \left[A^0\right]^T x \quad (12)$$

as,

accordingly, by integrating Eq. (13) into Eq. (12) and displacing

$$\sum_{l=1}^{L}\varsigma_l\left(\left[A^l\right]^T x - \left[b^l\right]\right) \to \sum_{l=1}^{L}\left|\left[A^l\right]^T x - \left[b^l\right]\right| \quad (13)$$

inequality sides,

$$\left[A^0\right]^T x + \sum_{l=1}^{L}\left|\left[A^l\right]^T x - \left[b^l\right]^T\right| \leq \left[b^0\right] \quad (14)$$

finally, Eq. (14) is linearized using axillary parameters Z and W, and the uncertainty budget $\gamma$ is applied to be less conservative as [40],

$$\sum_{l=1}^{L}|Z_l| + \gamma.\max|W_l| + \left[A^0\right]^T x \leq b^0 \quad (15)$$

where [40],

$$Z_l + W_l = b^l - \left[A^l\right]^T x, l = 1, 2, ..., L \quad (16)$$

Generally, objective function parameters are uncertain, so they are modeled by auxiliary variables. If matrix c is uncertain, by using as an auxiliary parameter, the robust counterpart of Eq. (3) can be written as follows [40],

$$O^{RO} = \min \Gamma^O \quad (17)$$

similar to Eq. (15), by introducing auxiliary parameters $Z'$, $W'$ for uncertain parameter c, and considering budget $\gamma'$, Eq. (17) is converted as follows,

$$\Gamma^O \geq c'x + b + \gamma' \cdot \max|W_l'| + \sum_{l=1}^{L}|Z_l'| \quad (18)$$

$$Z' + W' = \Delta O \quad (19)$$

Eq. (18) can be linearized, for instance, absolute value W can be linearized as follows,

$$\max |W| \leq \alpha \quad (20)$$

$$-\alpha \leq W \leq \alpha \quad (21)$$

## IV. CASE STUDY FORMULATION

To apply DRPs, it is assumed that the utility suggests two options for consumers. First, offering consumers desired AC temperature setpoints to adjust their indoor temperature. Second, shifting selected appliance operations from peak hours to off-peak hours. Consumers who participate in DRPs receive financial rewards or face penalties. In this study, four cases are developed to examine the impact of uncertainty in demand and occupancy data and application of DRPs.

- Case (a): Without consideration for uncertainties and no application of DRPs.
- Case (b): Consideration for uncertainty in data and no application of DRPs.
- Case (c): Without consideration for uncertainties and application of DRPs.
- Case (d): Consideration for uncertainties and application of DRPs.

### A. Case (a)

As all data are assumed deterministic and there is no DRP contract between consumers and the utility, the consumers energy cost is as follows,

$$\text{Cos} t : \sum_{h \in H} c_h^u p_h^u \quad (22)$$

### B. Case (b)

As this case examines the impact of uncertainties in data for demand and occupancy levels and no application of DRPs, based on the methodology described in the robust counterpart optimization section, Eq. (22) can be written as an uncertain optimization problem given by,

$$O_{case\,(b)} = \min \left\{ \sum_{h \in H} c_h^u p_h^d \right\} \quad (23)$$

According to Eqs. (17)-(21), by applying auxiliary parameter $\Gamma^1$, Eq. (23) can be rewritten as follows,

$$O_{case\,(b)}^{RO} = \min\{\Gamma^1\} \quad (24)$$

where $O_{case(b)}^{RO}$ is robust counterpart of $O_{case(b)}$ subject to,

$$\Gamma^1 \geq \sum_{h \in H} c_h^u p_h^d \quad (25)$$

$$\forall_h \in H, Z_h^{p^d} = \left\{ \varsigma^{p^d} \in R^H : \left\|\varsigma^{p^d}\right\|_\infty \leq 1, \left\|\varsigma^{p^d}\right\|_1 \leq \gamma_h^{p^d} \right\} \quad (26)$$

$$\Gamma^1 \geq \sum_{h \in H} c_h^u p_h^d + \gamma_{h,case\,(b)}^{p^d} \cdot \max\left(\left|W_{h,case\,(b)}^{p^d}\right|\right) + \sum_{h \in H}\left|Z_{h,case\,(b)}^{p^d}\right| \quad (27)$$

$$W_h^{p^d} + Z_h^{p^d} = -\Delta p_h^d \cdot c_h^u \quad (28)$$

Eq. (27) can be easily linearized by using Eqs. (20) and (21). Also, a power balance constraint is required to model the total internal load,

$$p_h^u = p_h^d, \forall_h \in H \quad (29)$$

Since consideration for uncertainties increases the margin of safety, the power balance equality in Eq. (29) can be expressed as [40],

$$p_h^u \geq p_h^d, \forall_h \in H \quad (30)$$

since consumer demand contains occupancy levels, Eq. (2) can be rewritten as,

$$p_h^d = p_h^{d'} + p_h^{AC} \quad (31)$$

$$p_h^{d'} = +p_h^{shift} + p_h^{n-shift} + p_h^{mis} \quad (32)$$

By combining Eqs. (1) and (2) with Eqs. (31) and (32), power balance constraint can be rewritten as,

$$p_h^u \geq \sum_{k \in K}\left[-\alpha_k \times p_{(h-k)}^{AC} + \beta_k \times Occ_{(h-k+1)}\right] + p_h^{d'}, \forall_h \in H \quad (33)$$

According to Eqs. (14)-(16), robust counterpart of Eq. (33) can be derived as,

$$p_h^u \geq \sum_{k \in K}\left[-\alpha_k \times p_{(h-k)}^{AC} + \beta_k \times Occ_{(h-k+1)}\right] + p_h^{d'}$$
$$+ \gamma_{h,case(b)}^{p^{d'}} \cdot \max\left(\left|W_{h,case(b)}^{p^{d'}}\right|\right) + \left|Z_{h,case(b)}^{p^{d'}}\right| \quad (34)$$
$$+ \gamma_{h,case(b)}^{occ} \cdot \max\left(\left|W_{h,case(b)}^{occ}\right|\right) + \left|Z_{h,case(b)}^{occ}\right|$$

$$\forall_h \in H : Z_{h,case(b)}^{p^{d'}} + W_{h,case(b)}^{p^{d'}} = \Delta p_h^{d'} \quad (35)$$

$$\forall_h \in H : Z_{h,case(b)}^{occ} + W_{h,case(b)}^{occ} = \sum_{k \in K} \beta_k \times \Delta Occ_{(h-k+1)} \quad (36)$$

Notice that Eq. (34) is linearized based on Eqs. (20) and (21).

### C. Case (c)

In this case, consumers participate in DRPs without consideration for uncertainties in data, and a multi-objective optimization is developed.

#### 1. Objective functions

The problem is formulated as a multi-objective optimization where (a) electricity load demand reduction, (b) adjustment in occupants thermal comfort AC temperature setpoints, and (c) minimizing consumer cost are achieved subject to the related constraints.

*(a) Load demand reduction objective function*

Typically, during peak hours the HEMS requires consumers to match their load demand with the expected load demand and minimize the difference between them at time h. It can be accomplished by adjusting the AC setpoint or shifting loads from on-peak to off-peak hours. In addition, when consumer load demand is less than desired demand, the corresponding objective function takes a value of zero. The first objective function is defined as [41],

$$O_{case\,(c)}^1 = \min \sum_{h \in H}\left|p_h^d - p_h^{d,des}\right| \quad (37)$$

*(b) Occupants thermal comfort objective function*

This objective function relates to the utility DRP, which may jeopardize occupants thermal comfort by adjusting the AC temperature setpoints. So, to minimize dissatisfaction regarding the AC temperature setpoints, the difference between the desired and indoor temperature has to be minimized. The occupant thermal comfort objective function is expressed as follows,

$$O^2_{case\,(c)} = \min \sum_{h \in H}\left(T_h^{AC,set} - T^{AC,des}\right).Occ_h \quad (38)$$

*(c) Consumer cost function*

Participation in DRPs must reduce customer costs. If consumers decrease the load demand relative to the utility desired load demand, they earn incentive fees. Otherwise, they will be monetarily penalized. This has been expressed as the third objective function given by,

$$O^3_{case\,(c)} = \min \sum_{h \in H} c_h^u p_h^d + c_c\left(p_h^d - p_h^{d,des}\right) \quad (39)$$

*2. Constraints*

*a) Demand response constraints*

It follows that if shiftable appliances are switched on at time h:

- In the previous steps, they must be off.
- They must remain on for the required time to accomplish their cycle.
- Once the cycle is accomplished, it must be switched off.

These constraints are formulated in Eqs. (40) and (41). Due to these constraints, the demand can be moved to the next hours if there is enough time left to perform the cycle in the dispatch period [43]. Eqs. (42) and (43) states that appliances cannot start before the predetermined start time and must complete their processes before the predetermined end time. Moreover, if the demand shift is planned, the entire load will be shifted to the following time step [41].

$$\forall_s \in S : \sum_{h \in H} u_{s,h} = N_s \quad (40)$$

$$\forall_s \in S, \forall_h \in H : u_{s,h+1} \geq \frac{u_{s,h}}{N_s}(N_s - \sum_{\tau=1}^{h} u_{s.\tau}) \quad (41)$$

$$\forall_s \in S : \sum_{h=1}^{H_s^{start}-1} u_{s,h} = 0 \quad (42)$$

$$\forall_s \in S : \sum_{h=H_s^{end}+1}^{H} u_{s,h} = 0 \quad (43)$$

*(a) Power balance constraint*

This constraint is described in case (b) as Eq. (29).

*(b) AC constraints*

The optimization problem may determine some uncomfortable temperatures for residents. Thus, we have considered the desired AC temperature setpoints to be 23.33°C for residences, and the AC temperature setpoints should not exceed 28.55°C at any point in time by utility [11],

$$\forall_h \in H : \left|T_h^{AC,set} - T^{AC,des}\right| \leq 5.22 \quad (44)$$

It is also possible that the optimization problem keeps the AC temperature setpoints at their maximum value for all hours (Eq. 44), which is not desired. Therefore, we need to limit the total variations of the AC temperature setpoints against the desired AC setpoints. This constraint is expressed by Eq. (45), where 19.44°C is chosen according to [11].

$$\sum_{h=1}^{H}\left|T_h^{AC,set} - T^{AC,des}\right| \leq 19.44 \quad (45)$$

Furthermore, to prohibit the overcooling in the summer, suppose the AC temperature setpoints will never be less than the desired temperature. This constraint is expressed as follows [11],

$$\forall_h \in H : T_h^{AC,set} \geq T^{AC,des} \quad (46)$$

*D. Case (d)*

As described, the robust equivalent of the multi-objective function expressed in case (c) can be calculated. Since the second objective function is nonlinear, it can be linearized using an auxiliary variable $y_h$, as shown in Eqs. (20) and (21). In addition, the second and third objective functions include uncertain variables. So, the auxiliary variable models their uncertainty [40]. So, Eq. (38) can be rewritten as Eq. (47),

$$O^{1,Ro}_{case\,(d)} = \min \sum_{h \in H} y_h \quad (47)$$

where,

$$\forall_h \in h : y_h \geq \left|p_h^d - p_h^{d,des}\right| \quad (48)$$

Since all objective functions contain uncertain variables, auxiliary variables $\Gamma^1_{case\,(d)}, \Gamma^2_{case\,(d)}$, and $\Gamma^3_{case\,(d)}$ are used to obtain their robust counterpart.

$$O^{1,Ro}_{case\,(d)} = \min \Gamma^1_{case\,(d)} \quad (49)$$

$$O^{2,Ro}_{case\,(d)} = \min \Gamma^2_{case\,(d)} \quad (50)$$

$$O^{3,Ro}_{case\,(d)} = \min \Gamma^3_{case\,(d)} \quad (51)$$

$$\Gamma^1_{case\,(d)} \geq \sum_{h \in H} y_h \quad (52)$$

$$\Gamma^2_{case\,(d)} \geq \sum_{h \in H}(T^{AC,set} - T^{AC,des}).Occ_h \quad (53)$$

$$\Gamma^3_{case\,(d)} \geq \sum_{h \in H}(c_h^u p_h^d + c_c(p_h^d - p_h^{d,des})) \quad (54)$$

According to Eqs. (18) and (19), robust counterpart of Eqs. (52)-(54) can be derived as follows by including $\gamma_h^{occ}$ and $\gamma_h^d$ as occupancy level and demand budgets, respectively, and $W_h^{occ}$, $Z_h^{occ}$, $W_h^d$, $w_h^{d'}$, $Z_h^{d'}$, and $Z_h^d$ as auxiliary variables.

$$\Gamma^2_{case\,(d)} \geq \sum_{h \in H}(T^{AC,set} - T^{AC,des}).Occ_h +$$
$$\gamma_h^{occ}.\max(|w_h^{occ}|) + \sum_{h \in H}|Z_h^{occ}| \quad (55)$$

$$\Gamma^1_{case\,(d)} \geq \sum_{h \in H} y_h + \gamma_h^d.\max(|w_h^d|) + \sum_{h \in H}|Z_h^d| \quad (56)$$

$$\Gamma^3_{case\,(d)} \geq \sum_{h \in H}(c_h^u p_h^d + c_c(p_h^d - p_h^{d,des}))$$
$$+\gamma_h^d.\max(|w_h^{d'}|) + \sum_{h \in H}|Z_h^{d'}| \quad (57)$$

$$\forall_h \in H, w_h^{occ} + Z_h^{occ} = -d(Occ_h)\Delta(T^{AC,set} - T^{AC,des}) \quad (58)$$

$$\forall_h \in H, w_h^d + Z_h^d = -d(p_h^d) \quad (59)$$

$$\forall_h \in H, w_h^{d'} + Z_h^{d'} = -d(c_h^{tu} p_h^d) \quad (60)$$

where perturbation set for each uncertain variable is as follows,

$$Z_h^d = \left\{\varsigma^{p^d} \in R^H : \left\|\varsigma^{p^d}\right\|_\infty \le 1, \left\|\varsigma^{p^d}\right\|_1 \le \gamma_h^{p^d}\right\}$$

$$Z_h^{occ} = \left\{\varsigma^{occ} \in R^H : \left\|\varsigma^{occ}\right\|_\infty \le 1, \left\|\varsigma^{occ}\right\|_1 \le \gamma_h^{occ}\right\}$$

For simplicity of illustration, all constraints for case (d) are included in cases (b) and (c).

*E. Multi-objective genetic algorithm*

Multi-objective GA is a particular type of evolutionary algorithm. In nature, better generations are produced from better chromosome combinations. In GA, several answers to the problem are generated at first. Then, the genetic algorithm selects the chromosomes with good features (for example, intelligence), combines them, and makes a mutation. Finally, the current population with the new population from combination and mutation are merged. If these features are inherited, the number of intelligent people will be higher in the next generation. The community will become smarter as this trend continues. Thus, multi-objective GAs can continuously improve each generation in terms of different characteristics by using a straightforward method [42, 43].

## V. RESULTS AND DISCUSSION

*A. Assumption and data collection*

The following assumptions and approaches are considered in this study for numerical analysis.

- The numbers of the Wi-Fi connection are derived from the history of the Wi-Fi connection hourly from Jun 10th, 2019 through Oct 7th, 2019. Also, electrical consumption is gathered from the local utility [44].
- There are usually three people in this house, but it varies from time to time [44].
- Washing machines and dishwashers are considered shiftable loads.
- It is assumed that indoor temperature equals ambient temperature.
- Forecasting and simulation are carried out for 12 hours. (It is applicable for larger periods if needed).
- For all uncertain parameters, 0.9 and 1.1 times their nominal value are considered the box of uncertainty [41].
- Without losses of generality, each time slot is taken to equal 1 hour, which means power demand in kW is the same as energy consumption in kWh.
- The financial fines and rewards are assumed to be the same and remain constant throughout all hours [41].
- It is assumed the house is demand responsive through HEMS.
- For forecasting and data preparation, Python is used and MATLAB is employed to solve the multi-objective problem.

*B. Machine learning results*

Neural networks are suitable for data with a very high number of features, such as image processing [45], [46]. For data with very few features the decision tree outperforms [35]. The parametric values for regression and classification process with MLP-ANN are given in Table I and II. In this study, occupancy samples are heavily imbalanced. That is, the majority of train data belong to the first class, i.e., resident presence, and the minority belongs to the second class, i.e., resident absence. In this case, the MLP-ANN classification process faces difficulties when learning the minority class. To illustrate this issue, the MLP-ANN is utilized for the classification process and the second-class precision, recall, and f1-score parameters are zero. That means the classification process is not suitable. So, regression is used. The parametric values for random forest and lightGBM algorithms are listed in Table III. As shown in Table IV, MSE, RMSE, and MAE are 0.97, 0.99, and 0.66 for MLP-ANN regression. According to Table IV, using lightGBM improves the MSE, RMSE, and MAE evaluation metrics by 3.52, 1.75, and 5.47 percent, respectively, and using the random forest algorithm improves them by 42.18, 23.94, and 40.6 percent. Therefore, random forest outperforms MLP-ANN regression and lightGBM. Finally, outcomes of the normalized random forest algorithm are used in the following simulations. Forecasting results with the random forest algorithm are illustrated in Fig. 1. This graph shows a strong correlation between electricity demand and occupancy levels, indicating that the forecasting is highly accurate. Therefore, we can conclude that for the imbalanced dataset, random forest outperforms. Other results are presented in the following cases.

*1. Case (a)*

This case is considered as the benchmark, without consideration for uncertainties and no application of DRPs. The total consumer cost is 6.942 $ for 12 hours.

*2. Case (b)*

The robust counterpart optimization along with uncertainty budgets is applied here. The budgets have transformed the cost function of the first case into an optimization problem. Each uncertain variable has 13 budgets. Since demand and occupancy levels are uncertain, there are 169 states on the whole. Due to a large number of states, we only examine points where both parameters have the same budget. Consumer total costs in this case are shown in Fig. 2. According to Fig. 2, when both budgets are zero, consumer total costs are equal to cost of case (a). That is, a zero budget indicates that the parameters are at their nominal (certain) values. However, as budget uncertainty grows, consumer expenses increase. For instance, consumer total costs are increased by 36 percent when uncertainty is considered at all hours. Additionally, robust counterparts are less conservative when budgets are growing. In the case that both uncertain budgets change from 0 to 1, robust counterpart optimization allocates these budgets to demand and occupancy in the hours that the cost is minimum i.e., the lowest electricity rate hours. As the budget increases, there is more uncertainty during high electricity price hours. Accordingly, the occupancy and demand for the highest electricity rate will be high when budgets go from 11 to 12. Numerically, by changing both budgets from 0 to 1 and 11 to 12, consumer costs rise by 0.08 and 5.37 percent, respectively.

TABLE I
Parametric values for MLP-ANN regression and classification

| Regression parameter | Value | Classification parameter |
|---|---|---|
| Solver | Adam | Solver |
| Alpha | 0.0001 | Alpha |
| Batch_size | Auto | Batch_size |
| Learning_rate | 0.001 | Learning_rate |
| Power_t | 0.5 | Activation function |
| Max_iter | 4000 | Max_iter |
| Shuffle | True | Shuffle |
| Random state | 42 | Epochs |
| Hyperparameter optimizer | Bayesian | Lost function |

TABLE II
Parametric values for MLP-ANN

| Layers | Number of nudes | Activation function |
|---|---|---|
| Input | 15 | - |
| Hidden 1 | 10 | Relu |
| Hidden 2 | 5 | Relu |
| Output | 1 | sigmoid |

TABLE IV
Results: MSE, RMSE and MAE for MLP-ANN, random forest, and lightGBM algorithms

| Evaluation metric | MSE | RMSE | MAE |
|---|---|---|---|
| Random forest | 0.56307 | 0.75038 | 0.39525 |
| LightGBm | 0.93923 | 0.96914 | 0.62905 |
| MLP-ANN | 0.97357 | 0.98645 | 0.66541 |

*3. Case (c)*

In this case, consumers participate in DRPs without consideration for uncertainties. The multi-objective GA is employed to solve the optimization problem, and its convergence curve is depicted in Fig. 3. Table V shows the transferred demand for shiftable loads at various hour intervals. According to Table V, shiftable loads are transferred from the hot hours of the day, to the nighttime hours, when cooling loads are less prevalent. Demand transfers arise from the multi-objective function, which aims at minimizing consumer costs and surplus demand while avoiding fines. AC temperature setpoints are shown in Fig. 4. As shown in Fig. 4, the AC temperature setpoint equals the desired temperature, which is 23.85 for most hours. This is due to the occupants thermal comfort objective function, which attempts to keep the indoor temperature as close to desired as possible. Furthermore, consumer total cost is 6.025 $, which is improved by 13.2

TABLE III
Parametric values for random forest and LightGBM

| Random forest parameter | Value | LightGBM parameter | Value |
|---|---|---|---|
| Criterion | Gini | Boosting type | Gbdt |
| Random-state | 0 | Random state | 0 |
| N_estimators | 400 | N_estimators | 400 |
| Min_impurity_decrease | 0 | Num-leaves | 31 |
| Max_depth | None | Max_depth | None |
| Bootstrap | True | Objective | Regression |
| Random state | 0 | Learning-rate | Callbacks |
| Max_features | Auto, Sqrt, Log2 | Subsanple-for-bin | 20000 |

percent in comparison with case (a). This improvement can be attributed to the load demand reduction objective function. Consumers try to bring their demands closer to the desired demand. This demand reduction is possible by shifting loads to lower-rate hours and increasing the AC temperature setpoints. It should be noted that the occupant thermal comfort objective function causes the AC to run at low temperatures (close to desired) during the majority of the hours, which increases costs and demand.

*4. Case (d)*

The concurrent effect of considering uncertainties and application of DRPs are investigated in this case. Also, the multi-objective GA convergence curve is similar to Fig. 3. According to Fig. 4, the AC temperature setpoints are the same in uncertain and certain states for most hours. Only in the late hours of the night, when the electricity rate is lower, the AC setpoints decrease due to the uncertainty. This minor difference is due to the occupancy level coefficient in the occupant thermal comfort objective function. Because the difference between the two temperatures is small, and any change in occupancy levels caused by uncertainty, significantly increases the occupant thermal comfort objective function. Thus, the robust optimization aims to keep the AC temperature setpoints as close to the desired setpoints as possible. In addition, uncertainty has impact on shiftable load performance. According to Table VI, like case (c), the washing machine operation is moved from 17:00 and 18:00 to 21:00 and 22:00 o'clock, respectively. However, the dishwasher continues to run at 13:00 and only changes from 20:00 to 22:00. This change is because the hourly price determines the decision to transfer loads between different hours. Since, in this study, the hourly electric rate is assumed to be deterministic, in both cases, the optimization problem reached the nearly same schedule. In addition, uncertainty decreases the AC temperature setpoints at 21:00 to 22:00 (see Fig. 4).

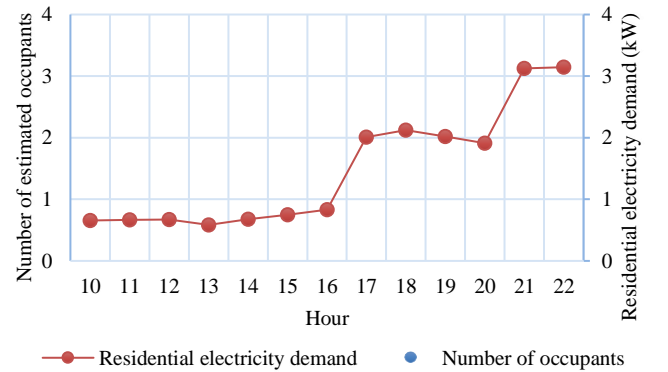

Fig. 1. Forecasted occupants and residential electricity demand for 12 hours.

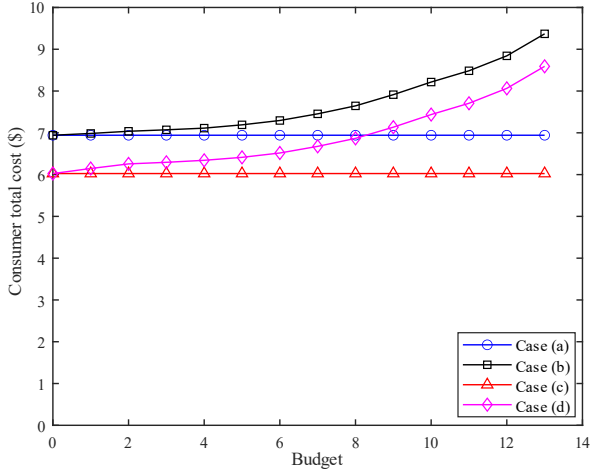

Fig. 2. Consumer total cost in different cases. where the horizontal axis indicates occupancy and demand budgets, for example, number 3 represents that occupancy and demand budgets are both 3.

Consequently, demand has increased during these hours So, the optimization problem prevents appliances from being transferred during these hours to avoid extra costs. Based on Fig. 2, the total cost of consumers in case (c) is equal to the total cost of consumers with no budgets in this case. Also, in the worst condition of case (d), i.e., uncertainty in all hours, the consumer costs improve by 9% compared to the worst point in case (b), which is due to the impact of DRPs.

In short, as shown in case (b), taking uncertainty into account raises consumer expenses. Therefore, according to [44] the goal of considering uncertainty is not only to achieve a better solution but also to ensure that the solution satisfies the constraints and remains feasible. But, implementing DRPs alleviates these costs (case(c)). So, uncertainty and DRPs have the opposite effect. Because the choice of uncertain budgets and DRPs depends on the financial and technical issues, Fig. 2 shows all the possible planning scenarios. Using the costs of the case (a) as a benchmark, implementing DRPs, and taking uncertainty into account up to 8 out of 12 hours is cost-effective. For example, in case (d), when the data is uncertain for 8 hours, the consumer costs are $6.94 which is nearly equal to the costs of case (a). However, in case (d) where there exists

TABLE V
Results: Transferred demand for shift-able loads in Case (c)

| From hour | To hour | Transferred demand (kW) |
|---|---|---|
| 13 | 22 | 0.5 |
| 20 | 22 | 0.5 |
| 17 | 21 | 1.5 |
| 18 | 22 | 1.5 |

uncertainty for more than 8 hours, it may be worthwhile to accept higher costs. Because the reliability of a system can be increased by using a more robust solution and the likelihood of unexpected failure can be reduced. Furthermore, it is less likely to have uncertainty in more than 8 hours out of 12. However, if we take the worst-case scenario approach, the consumer costs will be $8.59. So, robust counterpart optimization and budgets make sense and consumer costs decrease by 19%.

## VI. CONCLUSION AND RECOMMENDATIONS

This work proposed a privacy-aware method for predicting the occupancy level using the next-generation of machine learning algorithms. Data estimation results using machine learning tools demonstrated that MLP-ANN classification is ineffective for occupancy detection with non-uniform datasets. Furthermore, random forest outperforms MLP-ANN and lightGBM. Constraints can be violated if uncertainties are not considered, which may make the problem non-optimal and even infeasible. Because it is highly unlikely that all uncertain data will be in the worst condition at all times, robust counterpart optimization with budgeting was used. On the other hand, considering the uncertainty increases the costs and demand. So, DRPs were applied to reduce the demand and cost. The utility can adjust consumer AC temperature setpoints by implementing DRPs. However, utilities should not interrupt consumers' comfort by adjusting the AC temperature setpoints. Consequently, individuals' comfort was considered as an objective function. The multi-objective problem was solved in certain and uncertain states. The simulation results showed that considering the uncertainty with the budgets is more cost-effective than the worst-case scenario, and implementing DRPs can eliminate the deteriorating effects of uncertainty. As part of the future works, the following objects are proposed.

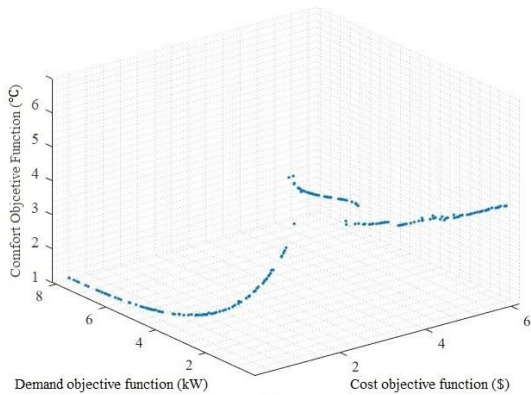

Fig. 3. Multi-objective GA convergence curve in Case (c).

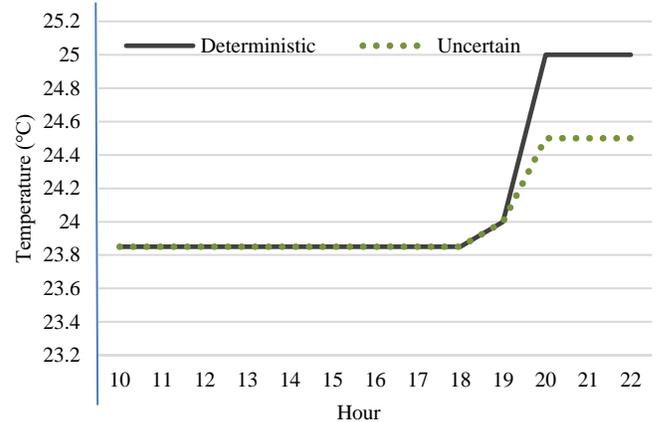

Fig. 4. AC set-point temperature in Cases (c) and (d).

TABLE VI
Results: Transferred demand for shift-able loads in Case (d)

| From hour | To hour | Transferred demand (kW) |
|---|---|---|
| 20 | 22 | 0.5 |
| 17 | 21 | 1.5 |
| 18 | 22 | 1.5 |

- Applying a smart city or multiple smart homes instead of a smart home.
- Formulating the problem from the investors or utility point of view.
- Considering uncertainties in other data such as market price.
- Considering comfort in other sections, namely dishwasher and washing machine operation time and finding the best function hours of these machines.